# Microstructure and superconducting properties of hot isostatically pressed MgB$_2$

T C Shields, K Kawano, D Holdom and J S Abell

School of Metallurgy and Materials, University of Birmingham, Birmingham, B15 2TT, UK.

**Abstract**
Bulk samples of MgB$_2$ have been formed by hot isostatic pressing (HIPping) of commercial powder at 100MPa and 950°C. The resulting material is 100% dense with a sharp superconducting transition at 37.5K. Microstructural studies have indicated the presence of small amounts of second phases within the material, namely MgO and B rich compositions, probably MgB$_4$. Magnetisation measurements performed at 20K have revealed values of J$_c$=1.3 x 10$^6$A/cm$^2$ at zero field, and 9.3 x 10$^5$A/cm$^2$ at 1T. Magneto optical (MO) studies have shown direct evidence for the superconducting homogeneity and strong intergranular current flow in the material.

## 1. Introduction
Since the discovery of superconductivity in MgB$_2$ at 39K [1] numerous attempts have been made to produce high density compacts. This has proved relatively difficult, mainly due to the high volatility of Mg at elevated temperatures. Despite this, MgB$_2$ compacts have been successfully prepared by various methods. These have included uniaxial pressing of mixtures of Mg and B [2-5], and sintering at high pressures, $\geq$ 3GPa, of MgB$_2$ powder [6-8]. The latter method has been the most successful in producing fully dense samples. HIPping has also been used to produce compacts. Nagamatsu et al [1] in their original studies pressed Mg and B powder at 700°C and 196MPa. Commercial MgB$_2$ powder has also been HIPped at 1000°C and 200MPa [9,10] to produce high density compacts. This paper reports the successful fabrication of MgB$_2$ compacts with high density and good superconducting properties by a HIPping process at 100Mpa and 950°C.

## 2. Experimental
Green compacts of commercial MgB$_2$ powder (Alfa Aesar 98%, particle size ~1µm) were pressed and placed in a mild steel can which was evacuated and sealed by welding. The initial diameter of the compacts were 13mm with a total length of 20mm, shrinking to 10mm diameter after HIPping. The HIPping machine used was a VG superhipper. The sample was heated to 950°C and held for 4hrs at a pressure of 100MPa, taking 1.5hrs to heat up and 1.5hrs to cool down. Ar gas was used to surround the can during the whole procedure. Samples were cut from the can for analysis by X–ray diffraction (XRD), scanning electron microscopy (SEM), AC susceptibility ($\chi$ac), vibrating sample magnetometry (VSM) and magneto-optical (MO) measurements. Details of the MO system are given elsewhere [11]. The density of the sample was measured as 2.63±0.02g/cm$^3$ which is close to the theoretical density of 2.625g/cm$^3$ [9].

## 3. Results

The XRD pattern of the cross-section of the compact is shown in Figure 1. The pattern shows primarily the $MgB_2$ phase with the presence of weak second phase peaks belonging to MgO and possibly $MgB_4$. An SEM micrograph of a polished cross-section of the sample is shown in Figure 2. The microstructure is dense and well connected with a grain size up to about 150µm. Energy dispersive X-ray analysis (EDX) was performed on Mg and Fe, but it was necessary to use wavelength dispersive X-ray analysis (WDX) for B and O because of their low atomic numbers. Analyses on the matrix phase ranged from Mg (37.1–38.2at%), B (56.6-60.2at%) and O (2.3 5.9at%). The Mg rich values and the detection of O in the matrix are probably due to finely dispersed MgO within the $MgB_2$ grains. A similar microstructure has been seen previously in samples with strong superconducting grains, which were thought to contain a nanomixture of $MgB_2$, MgO and a B–rich phase [2]. Transmission electron microscope (TEM) studies on sintered polycrystalline $MgB_2$ [12] have found MgO precipitates ranging in size from 10–500nm. It has been suggested that the mismatch with the $MgB_2$ matrix can be a source of dislocations. Smaller grains of darker contrast are also present (Figure 2) consisting of Mg (11.6-25.7at%) and B (77.6-88.2at%) and probably account for the $MgB_4$ peak in the XRD analysis. It should also be noted that the EDX results confirm no Fe contamination in the $MgB_2$ sample, in agreement with other workers [13].

The $\chi$ac transition is sharp as shown in Figure 3 with a $T_c$ onset of about 37.5K and $T_c$ endpoint of about 37K. Magnetisation loops were performed at 10K, 20K and 30K and are shown in Figure 4. The curve at 10K exhibits flux jumping which has already been seen and discussed for bulk $MgB_2$ [3]. Estimates of $J_c$ have been made at 20K and 30K using the Bean model equation [14], $J_c=30\Delta M/d$, where; $J_c$ is the critical current density ($A/cm^2$), $\Delta M$ is the hysteresis of the magnetisation curve ($emu/cm^3$) and d is the sample size (cm), in this case taken as the average sample dimension of 0.15cm. At 20K $J_c=1.3 \times 10^6$ $A/cm^2$ in zero field dropping to $9.3 \times 10^5 A/cm^2$ in 1T field. At 30K $J_c=5\times 10^5$ $A/cm^2$ in zero field dropping to $1 \times 10^5 A/cm^2$ in 1T field. These values are amongst the highest achieved for bulk $MgB_2$ and compare very favourably with those previously measured by magnetisation methods [2,5,6,9,10].

The excellent superconducting properties of the sample are also demonstrated by the MO images of Figures 5 and 6. The external magnetic fields were applied perpendicular to the sample surface and the bright regions indicate the presence of magnetic field. Figure 5 shows the results of the zero field cooling (ZFC) process, where the sample was cooled from above Tc to each temperature and the magnetic field applied. These images show the images for a field of 60mT. The images show no flux penetration into the sample up to and including 30K, indicating high Jc values of the sample at these temperatures. Flux is still excluded from the centre of the sample at 36K just below $T_c$ of the sample. Similarly Figure 6 shows images for the field cooling (FC) process, in which a field of 150mT was applied above $T_c$ before cooling to 5K. After cooling the magnetic field was set to 0T and the temperature gradually increased. There is extensive flux trapping and again no significant temperature dependence up to ~30K. The magnetic field is trapped uniformly by the sample indicating its superconducting homogeneity. Clearly the MO studies indicate well connected grains and strong inter-granular current flow, in the whole of the sample, up to temperatures of 30K, and in the centre of the sample up to $T_c$. These MO images demonstrate the excellent superconducting properties of these high density compacts, exhibiting superior flux exclusion and trapping behaviour to those previously reported by MO on bulk material [2,15].

## 4. Conclusions

High density samples of $MgB_2$ have been prepared from a commercial powder by a HIPping process at 100MPa and 950°C. XRD and SEM studies indicate the presence of small quantities of second phases of MgO and $MgB_4$. The samples have excellent superconducting properties with Tc=37.5K. $J_c$ values at 20K of $1.3 \times 10^6$ A/cm$^2$ at zero field and $9.3 \times 10^5$ A/cm$^2$ at 1T, have been estimated from VSM measurements. MO images indicate the superconducting homogeneity of the samples and the presence of strong intergranular current flow up to 30K.

## Acknowledgements

The authors acknowledge the following colleagues in the IRC in Materials Processing at the University of Birmingham for useful discussions on the HIPping process and the provision of facilities, Edwin Fasham, Mark Ashworth and Roy Huzzard.

**Figure Captions**

Figure 1. XRD pattern of the compact ($MgB_2$ indices are labelled). Steel peak is from the can that surrounds the sample.

Figure 2. SEM micrograph of polished cross-section.

Figure 3. $\chi$ac susceptibility transition of HIPped compact.

Figure 4. Magnetisation curves taken at 10K, 20K and 30K.

Figure 5. MO images at different temperatures for the ZFC process at a field of 60mT.

Figure 6. MO images of FC process for applied field of 150mT at various temperatures.

Figure 1

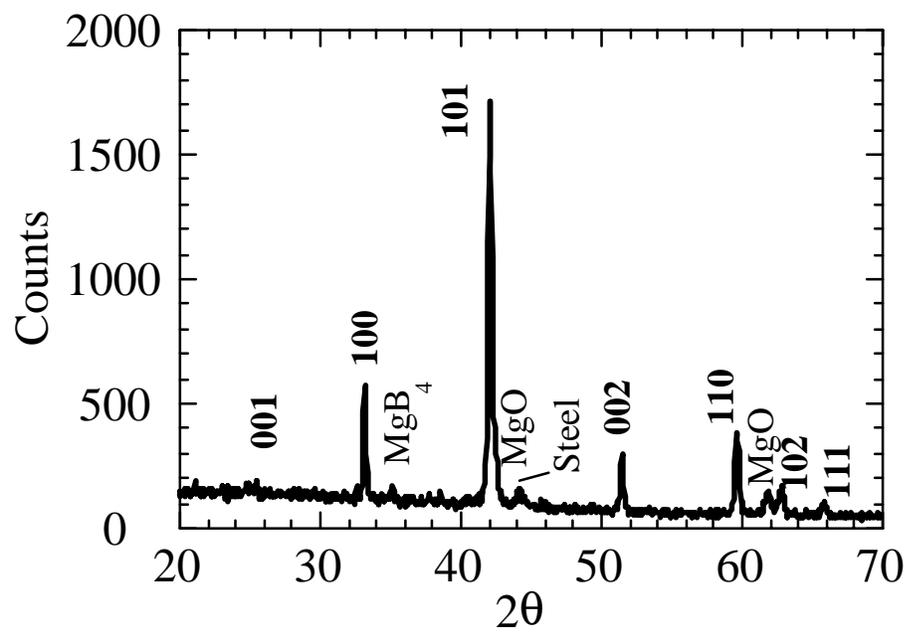

Figure 2

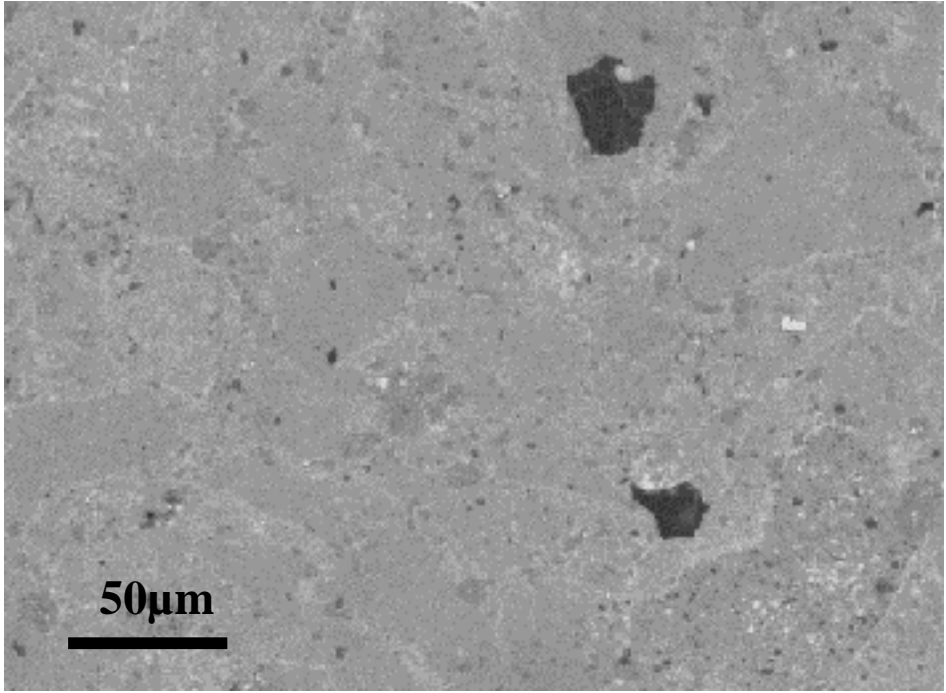

Figure 3

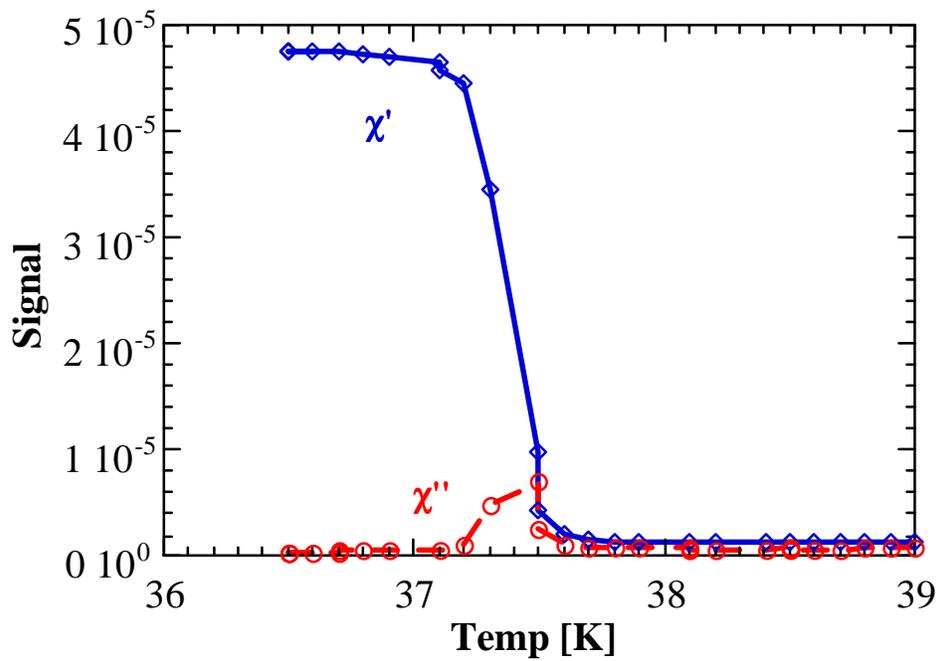

Figure 4

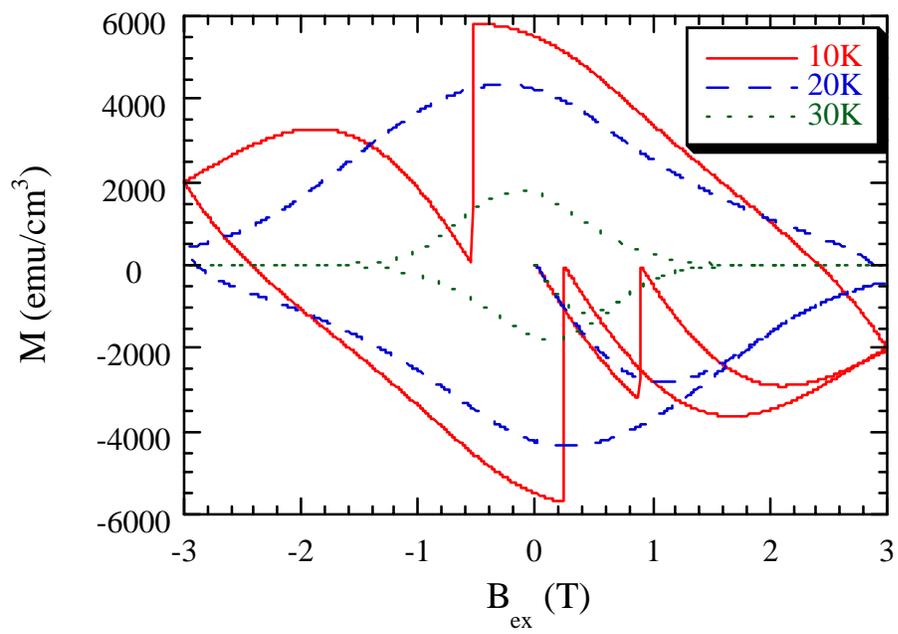

Figure 5

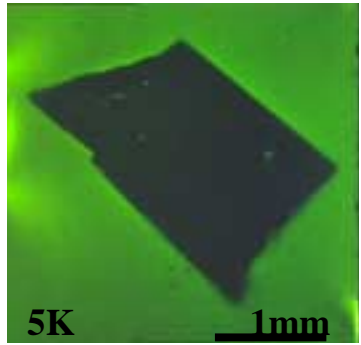
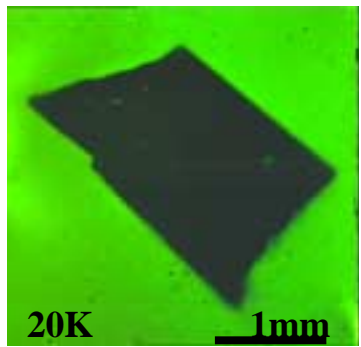
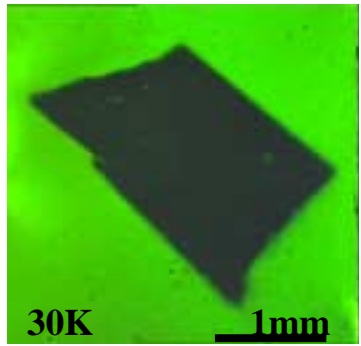
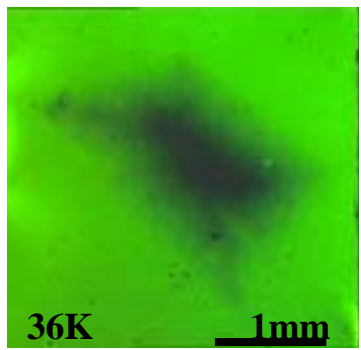

Figure 6

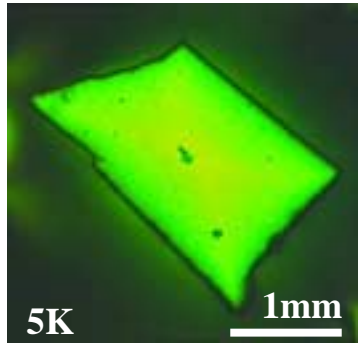 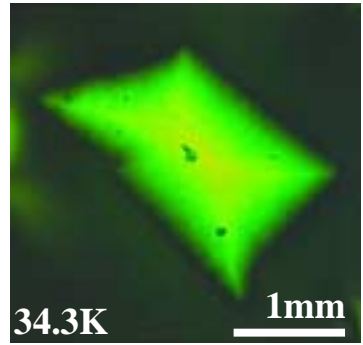
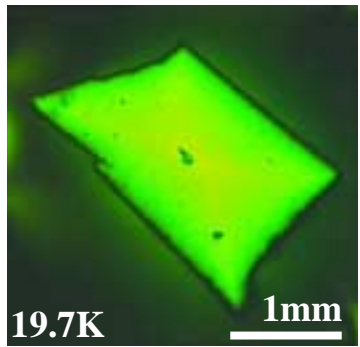 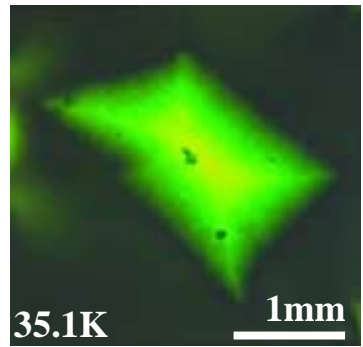
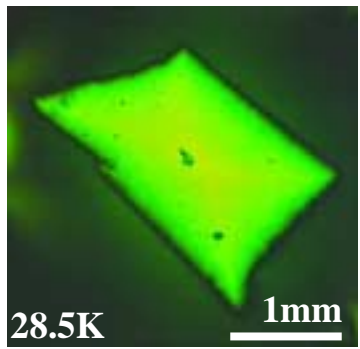 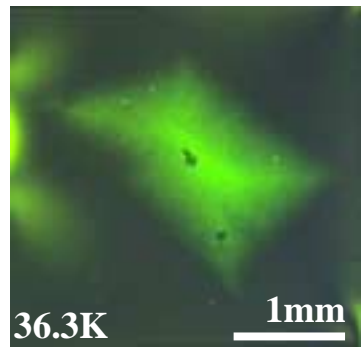